# Revisiting the third-order elastic constants of diamond: the higher-order effect

Mingqing Liao[1,2], Yong Liu[1,3], Yi Wang[2], Fei Zhou[1,4], Nan Qu[1], Tianyi Han[1], Danni Yang[1], Zhonghong Lai[1,5], Zi-Kui Liu[2], Jingchuan Zhu[1,3, *]

[1]School of Materials Science and Engineering, Harbin Institute of Technology, Harbin, Heilongjiang, 150001, China

[2]Department of Materials Science and Engineering, The Pennsylvania State University, University Park, PA, 16802, USA

[3]National Key Laboratory for Precision Hot Processing of Metals, Harbin Institute of Technology, Harbin 150001, Heilongjiang, China

[4]MIIT Key Laboratory of Critical Materials Technology for New Energy Conversion and Storage, School of Chemistry and Chemical Engineering, Harbin Institute of Technology, Harbin, Heilongjiang, 150001, China

[5]Centre of Analysis and Measurement, Harbin Institute of Technology, Harbin, Heilongjiang, 150001, China

**ABSTRACT**

In this letter, we study the higher-order effect on the third-order elastic constants (TOECs) of diamond using longitudinal stress-uniaxial strain (LSUS) approach based on density functional theory. The result shows that the higher-order effect on TOECs is not negligible in the shock wave experiments and similar calculations. By taking the higher-order elastic response (up to fifth order) into consideration, the convergence of TOECs against maximum stain gets improved significantly and resolves the discrepancy of several different theoretical methods and experiments.

---

* Corresponding author: fgms@hit.edu.cn (Jingchuan Zhu)

Owing to the excellent mechanical and thermal properties, diamond has been widely used in extreme physics. For example, the diamond anvil cell is used to create static pressure up to 400GPa[1,2] and the diamond structure can be kept stable at even 2TPa[3]. Besides, diamond may exhibit some unexpected properties, e.g. superconductivity[4], under extreme conditions. In the extreme condition, the higher-order elastic constants (HOECs) are of great help to understand the nonlinear behavior, such as thermal expansion and Grüneisen parameters[5], temperature and pressure dependence of the second-order elastic constant (SOECs)[6–8], ductility[9] and phase transformation[10]. However, even for the third-order elastic constants (TOECs), there exist large variations in diamond among different experiments and different calculations.

The TOECs of diamond was first reported by Grimsditch et al.[11] by combining experiments with Keating's theory[12] for diamond-structure, in which only the first two neighbors were taken into consideration. The results were subsequently improved by more accurate and general Keating's models[13,14]. To get rid of the empirical or experimental parameters, Nielsen[15] used the ab-initio method with three different strain modes to calculate the TOECs of diamond. To resolve huge variations among the above different results, Lang et al.[16] measured the TOECs of diamond by shock wave compression and revised the value by correcting an error in the pressure derivatives of SOECs[17]. However, there still exists a large difference between the experimental results and previous theoretical results. To further improve the predictive ability of theory, a new method that in analogous to the shock wave experiments is proposed and implied under the frame of density functional theory (DFT)[18] and molecular dynamics[19]. By such strategies, the agreements were improved when compared with experiments. Nevertheless, an apparent difference occurred in Hmiel's work between the strain-energy method (SEM) and longitudinal stress-uniaxial strain approach (LSUS), which was mainly ascribed to the different frequencies of various TOECs[18] and partially affected by HOECs (e.g. fourth-order elastic constants, FOECs). Recently, a new method separating each elastic constants (SEP) based on the numerical difference of the Piola-Kirchhoff stress was proposed[20], in which the results agreed well with the data of SEM in Hmiel's work[18] though the frequencies of each TOECs were different in those two methods. In addition, some recent experiments[21,22] are quite different from Lang's data[16] as well as Hmiel's LSUS data[18]. Hence, the higher-order effect must play a critical role in determining the TOECs of diamond.

In this work, we studied the HOECs effect on the measurements from shock wave experiments by first-principles calculations. All first-principles calculations based on the density functional theory (DFT) are performed using CASTEP[23] with ultra-soft pseudopotential[24], and Perdew-Burke-Ernzerhof (PBE) functional in the generalized gradient approximation (GGA)[25] framework. The cutoff energy and k-points are determined by the convergence test as 400 eV and 8×8×8, respectively. The unit cell with 8 atoms is used in the calculation and full relaxed before deformation. To eliminate the error caused by different symmetry[20], a tiny (~$10^{-10}$ lattice constants) random perturbation on atomic position is added to the initial structure. The cell is fixed for the deformed structure. The convergence criteria of relaxation for deformed-structure are $5×10^{-6}$ eV/atom in energy, 0.01 eV/Å in residual force per atom, $5×10^{-4}$ Å in max displacement, and 0.005 GPa in max stress. Workflow of calculating stress under different strains is implemented in Elastic3rd code[26].

For calculating TOECs in this work, we have employed the longitudinal stress-uniaxial strain (LSUS) method[18], which is an imitation of shock wave experiments[16]. In this method, the stress along a given direction can be expressed as **Eq(1)**.

$$\sigma'_1 = \rho_0/\rho \ [C'_{11}\eta'_1 + 1/2 C'_{111}\eta'^2_1 + O(\eta'^3_1)] \tag{1}$$

where $\rho_0$ and $\rho$ are the densities of the undeformed and deformed structures, respectively, and the superscript ′ indicates quantities in the local coordination system, rotating [100] to the given direction in the global coordination system, for more details, one is referred to the work by Lang et al.[27]), and the values are listed in **Table I** for [100], [110], and [111] directions.

**Table I** The strain mode and corresponding coefficients of longitudinal stress-uniaxial strain approach used in current work.

| Direction | $\sigma'_1$ | $C'_{11}$ | $C'_{111}$ | $\eta'_1$ | Strain modes |
|---|---|---|---|---|---|
| 100 | $\sigma_1$ | $C_{11}$ | $C_{111}$ | $\eta$ | $[\eta,0,0,0,0,0]$ |
| 110 | $\sigma_1+\sigma_6$ | $1/2(C_{11}+C_{12}+2C_{44})$ | $1/4(C_{111}+3C_{112}+12C_{155})$ | $\eta$ | $0.5[\eta,\eta,0,0,0,2\eta]$ |
| 111 | $\sigma_1+2\sigma_6$ | $1/3(C_{11}+2C_{12}+4C_{44})$ | $1/9(C_{111}+6C_{112}+2C_{123}+12C_{144}+24C_{155}+16C_{456})$ | $0.99\eta$ | $0.33[\eta,\eta,\eta,2\eta,2\eta,2\eta,]$ |

The strain-stress relationship along three different directions listed in **Table I** provides three independent equations for calculating TOECs and the pressure dependence of SOECs (as **Eq(2)** and a typo in Hmiel et al.[18] is fixed) gives another three equations. In this work, if not specifically noted, the SOECs and pressure derivative of SOECs are taken from Winey et al.[17] which is a revision of experimental data[28].

$$\begin{cases} C_{111} + 2C_{112} = -dC_{11}/dP(C_{11} + C_{12}) - C_{11} \\ C_{123} + 2C_{112} = -dC_{12}/dP(C_{11} + C_{12}) - C_{12} \\ C_{144} + 2C_{155} = -dC_{44}/dP(C_{11} + C_{12}) - C_{44} \end{cases} \quad (2)$$

**Fig. 1** shows the strain stress relationships along the [100], [110], and [111] directions, for strain ranging from -8% to 8%, obtained by the present work and from the previous works[15,22,29–34]. It shows that our result agrees well with experiments and the previous calculations. For the negative strain range, all the fittings are consistent with the experiments and DFT calculations. For the positive strain range, our fitting with $n=3$ using compression (negative) strain (line 1. Note that $n$ in the plots refers to the highest order elastic constants considered in the fitting, hereinafter) is quite close to Hmiel's LSUS result (line 4) for all three directions. Meanwhile, a deviation is observed when the fitting is compared with DFT calculations and experiments, which means the fitting with $n=3$ by compression strain only cannot reproduce the strain stress relationship for large positive strain. With $n$ increased to 4 (using three-order polynomial of $\eta'_1$ in **Eq.1**, shown as line 2 in **Fig. 1**), the deviation for the positive strain region is greatly reduced. With $n$ increased to 5 (line 3 in **Fig. 1**), the strain-stress relations from the DFT calculations for both negative and positive strains are well reproduced. These observations indicate that the mismatch between modeling with $n=3$ and DFT calculations is mainly due to the higher-order terms.

The strain amplitude has strong effects on the calculation of TOECs, and only when the maximum strain located in a small range, one can get the correct TOECs[35]. The TOECs of diamond vs the strain amplitude is illustrated in **Fig. 2** using the LSUS method. When modeled with $n=3$, there are small plateaus in $C_{144}$, $C_{155}$, and $C_{456}$, but not in $C_{111}$, $C_{112}$, and $C_{123}$. Since the SOECs and FOECs terms are odd, and TOECs term is even in **Eq.(1),** there is no improvement in obtaining TOECs for symmetrical amplitude when $n$ is increased to 4 ($n=4$ should improve the SOECs). However, when $n$ is increased to 5, all the TOECs converge in a large strain range. Hence, the fifth-order elastic constants (FFOECs) effect should be taken into consideration when the strain amplitude is larger than ~4%.

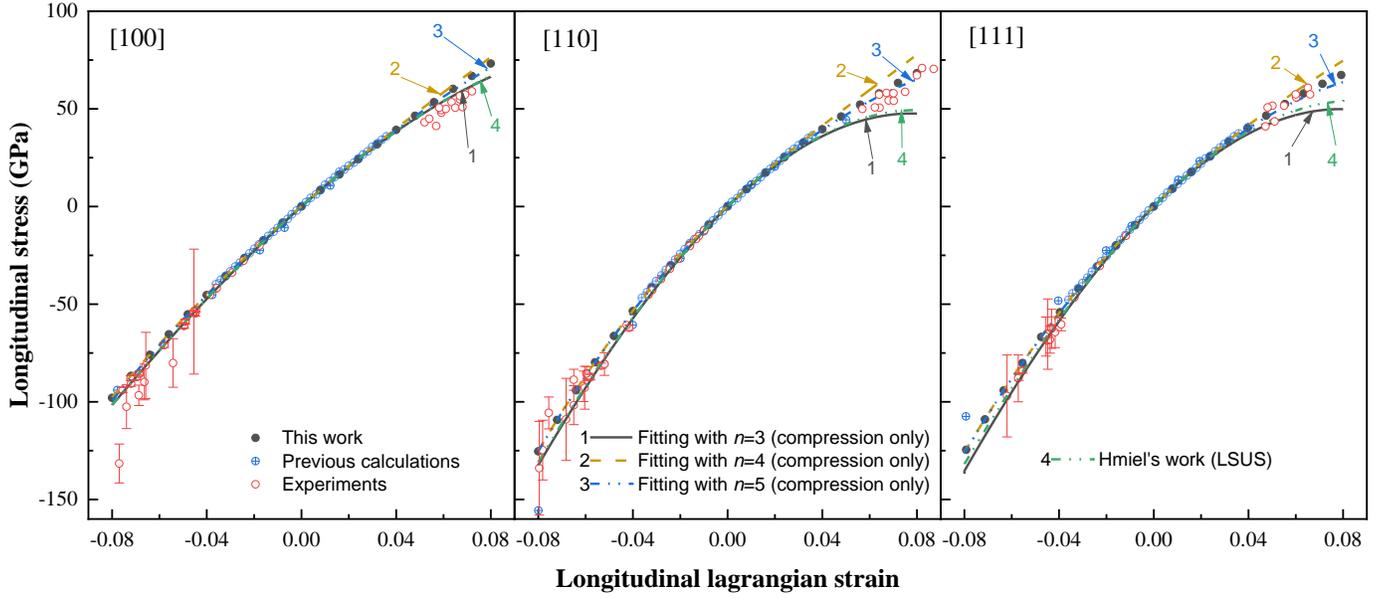

**Fig. 1** Longitudinal stress-strain relations along (a) [100], (b) [110], (c) [111] directions. The solid black circle is the data points calculated in current work by DFT, the open red circle is the experiments data points[29–34], and the cross blue circle is previous calculations[15,22]. The solid (line 1), dash (line 2) and dot (line 3) lines are the fitting results of compression strain ($\eta$ ranges from -0.08 to 0 with step of 0.008) with $n=3$, $n=4$ and $n=5$ ($n$ refers to the highest order of elastic constants taken into consideration in the fitting, hereinafter) respectively. The dash-dot-dot(line 4) line is calculated from **Eq(1)** with Hmiel's result[18] by the longitudinal stress-uniaxial strain approach.

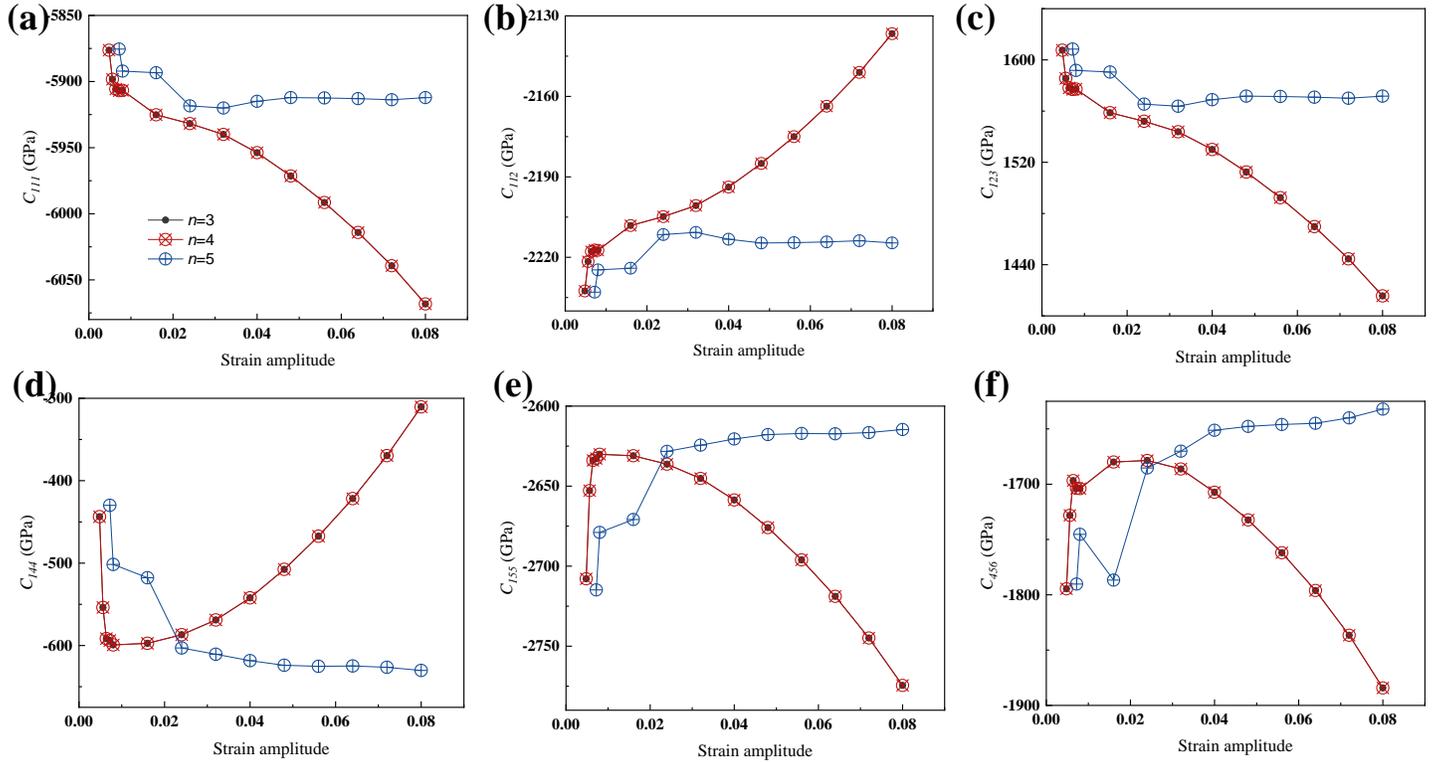

**Fig. 2** TOECs of diamond vs strain amplitude. Note: the strain step used in the fitting is 0.08% for strain range from 0.48% to 0.8% and 0.8% for strain range from 1.6% to 8%. (a) $C_{111}$; (b) $C_{112}$; (c) $C_{123}$; (d) $C_{144}$; (e) $C_{155}$; (f) $C_{456}$

Because both FOECs term and SOECs are odd, the effect of FOECs is totally included in SOECs when the strain amplitudes are symmetrical. However, the strain amplitudes adopted in experiments[16,21,22,34] and some calculations[15,17] are either compression only or tensile only. To investigate the effect of the asymmetrical strain amplitude on TOECs, **Fig. 3** shows the TOECs of diamond vs the central strain ($\eta_c$, the strain is symmetrical around the central strain which means the strain range is [$\eta_c-\eta_{max}$, $\eta_c+\eta_{max}$]) with different $n$. Our fitting at $\eta_c$=-4% agrees well with the shock wave experiments[16] except for $C_{456}$ due to the accumulated errors of other TOECs under the framework of the LSUS method. The TOECs monotonously increase ($C_{111}$, $C_{123}$, $C_{155}$ and $C_{456}$) or decrease ($C_{112}$ and $C_{144}$) with $\eta_c$ increasing from -4% to 4% with $n$=3. In this strain range, the varying amplitude of TOECs can reach ~3000GPa (e.g. from -7690GPa to -4700GPa for $C_{111}$), which is much larger than FFOECs effect (100-300GPa, as **Fig. 2**). With $n$=4, the behaviors of TOECs got improved a lot. Hence, the increase or decrease in the TOECs is mainly contributed by the FOECs effect, and such effect is quite obvious when the strain is not symmetrical with respect to zero. This is why some calculations[18,20] still give reasonable result though the FOECs is not taken into consideration. The fitted TOECs converge well for $\eta_c$ ranging from -4% to 4% when $n$ increases to 5.

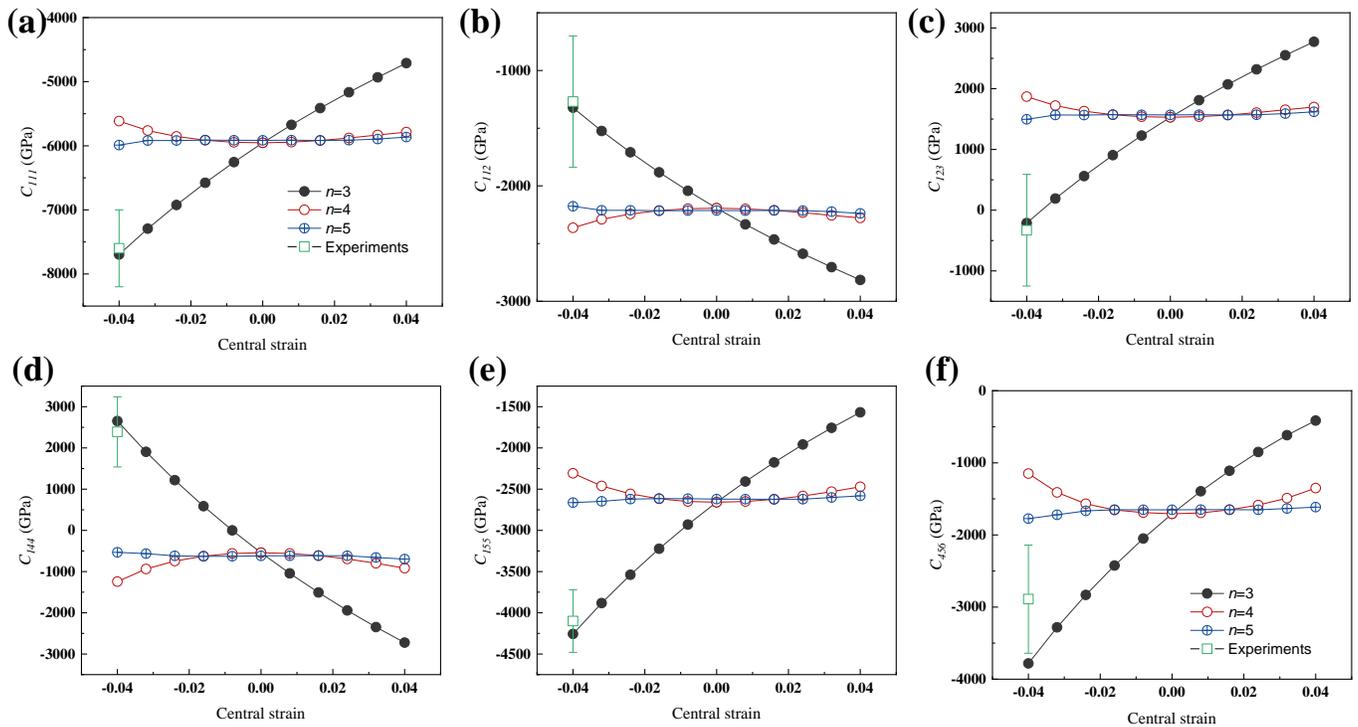

**Fig. 3** The TOECs of diamond vs central strain ($\eta_c$) with different $n$ and compared with previous result[18]. Note the strain range used in each point is [$\eta_c$-4%, $\eta_c$+4%] with the step of 0.8%. (a) $C_{111}$; (b) $C_{112}$; (c) $C_{123}$; (d) $C_{144}$; (e) $C_{155}$; (f) $C_{456}$

The full TOECs of diamond by LSUS under different conditions (different strain ranges, different $n$, and different SOECs and their derivative) are compared with previous works in **Table II**. The comparisons can be divided into three groups. The first group (including NO. 5, 6, 7, 13) uses the compression strain only and $n$=3. Hence, as analyzed above, the FOECs effect is included in the TOECs, causing such results inaccurate. In the second group (including NO. 3, 4, 8, 12, 18), all the parameters, including SOECs and their derivative, are obtained by first-principles calculations using symmetrical strain. Though different theoretical methods are used in this group, the TOECs agree quite well with each other. The rest (including NO. 1, 2, 9, 10, 11, 14, 15, 16, 17) can be grouped together. In this group, $C_{111}$, $C_{112}$, and $C_{155}$ are comparable, while $C_{123}$, $C_{144}$, and $C_{456}$ are dispersed even in the different experiments with the same method by the same authors[21,22]. Considering the large uncertainty in experiments, we recommend the second group as the true TOECs for diamond.

In summary, we have studied the TOECs of diamond and shows that the HOECs effect (up to fifth order) is significant both in experiments and calculations when the strain is not symmetrical with respect to zero. It shows that, by taking the higher-order effect into consideration, the TOECs convergence well. For the symmetrical strain case, the FOECs effect is totally included in SOECs and hence the values obtained in such condition give a reasonable evaluation of TOECs. Last, by calculating the TOECs in different conditions and compared to previous experiments and calculations, the difference among those works is resolved, and the recommended TOECs values for diamond are presented.

This work was supported by the China Scholarship Council and the China Postdoctoral Science Foundation also funded this project.

**Table II** TOECs of diamond by different methods. The strain range used in the current work is [-8%, 0] for LSUS-A, [-8%, 8%] for LSUS-S. Note: LSUS-A (-S) means LSUS with asymmetrical (symmetrical) strain amplitude and the SOECs and pressure derivative of SOECs in LSUS-S-PBE are taken from the PBE result in Ref.[18]. Unit in GPa.

| | Telichko et al.[21,22] | | | Nielsen[15] | Modak et al.[19] | Gupta et al[16–18] | | | Cousins[14] | Anastassakis et al.[11,13] | | Cao et al.[20] | Present work | | | | | |
|---|---|---|---|---|---|---|---|---|---|---|---|---|---|---|---|---|---|---|
| | Exp | Exp | SEM | SSM | MD[a] | Exp | LSUS | SEM | Exp +Theory | Exp+Theory | | SEP | LSUS-A ($n=3$) | LSUS-A ($n=4$) | LSUS-A ($n=5$) | LSUS-S ($n=3$) | LSUS-S ($n=5$) | LSUS-S-PBE ($n=5$) |
| $C_{111}$ | -7750±750 | -7660±500 | -7611 | -6300 | -7290 | -7600±600 | -7515 | -6026 | -6475 | -6260 | -7367 | -5876 | -7696 | -5615 | -5989 | -6068 | -5912 | -5912 |
| $C_{112}$ | -2220±500 | -1550±500 | -1637 | -800 | -866 | -1270±570 | -845 | -1643 | -1947 | -2260 | -2136 | -1593 | -1323 | -2363 | -2176 | -2137 | -2215 | -1646 |
| $C_{123}$ | 2100±200 | 3470±500 | 604 | 0 | -1191 | -330±920 | -960 | 606 | 982 | 112 | 1040 | 618 | -212 | 1869 | 1495 | 1416 | 1572 | 642 |
| $C_{144}$ | -1780±440 | -3130±300 | -199 | 0 | 243 | 2390±850 | 2693 | -200 | 115 | -674 | 186 | -197 | 2653 | -1243 | -533 | -310 | -630 | -241 |
| $C_{155}$ | -2800±220 | -2630±300 | -2799 | -2600 | -2996 | -4100±380 | -4223 | -2817 | -2998 | -2860 | -3292 | -2739 | -4256 | -2308 | -2663 | -2775 | -2615 | -2757 |
| $C_{456}$ | -30±150 | -700±300 | -1148 | -1300 | -4470 | -2890±750 | -2870 | -1168 | -135 | -823 | 76 | -1111 | -3782 | -1149 | -1774 | -1884 | -1632 | -1808 |
| NO. | 1 | 2 | 3 | 4 | 5 | 6 | 7 | 8 | 9 | 10 | 11 | 12 | 13 | 14 | 15 | 16 | 17 | 18 |

[a]MD: molecular dynamics. The result is revised according to Ref.[17].